\documentclass[10pt,aps,showpacs,twocolumn]{revtex4}
\usepackage{amssymb}

\usepackage{amsmath}
\usepackage{bm}
\usepackage{epsfig}

\def\be{\begin{equation}}
\def\ee{\end{equation}}
\def\bea{\begin{eqnarray}}
\def\eea{\end{eqnarray}}

\begin{document}

\title{{\Large Phase Controlled Continuously Entangled Two-Photon Laser with Double
}$\Lambda ${\Large \ Scheme}}
\author{C. H. Raymond Ooi}

\affiliation{{\sl Department of Physics, KAIST,
Guseong-dong, Yuseong-gu, Daejeon, 305-701 Korea\\
Max-Planck-Institut f\"{u}r Quantenoptik, D-85748, Garching,
Germany}}
\date{\today}

\begin{abstract}
We show that an absolute coherent phase of a laser can be used to manipulate
the entanglement of photon pairs of two-photon laser. Our focus is on the
generation of a continuous source of entangled photon pairs in the double $%
\Lambda $ (or Raman) scheme. We study the dependence of steady state
entanglement on the phase and laser parameters. We obtain a relationship
between entanglement and two-photon correlation. We derive conditions that
give steady state entanglement for the Raman-EIT scheme and use it to
identify region of steady state macroscopic entanglement. No entanglement is
found for the double resonant Raman scheme for any laser parameters.
\end{abstract}

\maketitle

\section{Introduction}

Entangled photon pairs is an integral asset to quantum communication
technology with continuous variables \cite{qtm comm}. A bright source of
entangled photon pairs could be useful also for quantum lithography \cite
{qtm litho}. Transient entanglement of a large number of photon pairs has
been shown to exist for cascade scheme \cite{Han},\cite{transient}, and
double Raman scheme \cite{Kiffner}. The transient regime does not provide a
continuous source of entangled photon pairs that could be as useful and
practical as typical lasers in c. w. operation. One might wonder whether the
entanglement still survives in the long time limit.

\begin{figure}[tbp]
\center\epsfxsize=8.0cm\epsffile{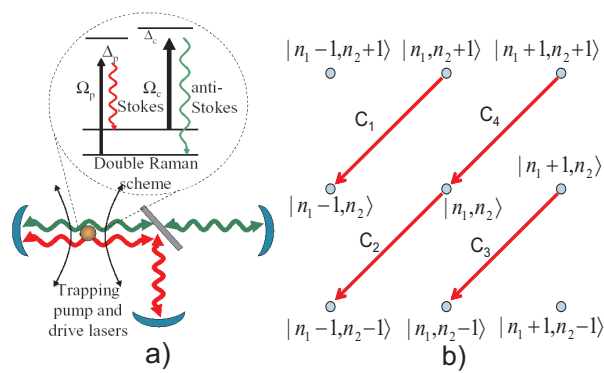}
\caption{a) Double Raman atom in doubly-resonant optical cavity. The atom is
a trapped by an optical dipole force and driven by a pump laser and a
control laser. \emph{Raman-EIT} (REIT) scheme ($\Omega _{c},\Delta
_{c}(=\Delta )>>\Omega _{p},\protect\gamma $) and \emph{double resonant Raman%
} (DRR) scheme ($\Omega _{c}=\Omega _{p},\Delta _{c}=\Delta _{p}=0$) would
be the focused for analysis. b) Four off-diagonal two-photon emission
density matrix elements with their respective coefficients $C_{k}$.}
\label{REDlaserscheme}
\end{figure}

In this paper, we use the coherent phase of the controlling lasers to
generate a large number (macroscopic) of entangled photon pairs in the
steady state. This also shows the possibility of coherently controlling
entanglement in the steady state. We focus on the Raman-EIT scheme (Fig. \ref
{REDlaserscheme}a) that has been shown to produce nonclassically correlated
photon pairs in single atom and many atoms cases.

First, we discuss the physics of a two-photon emission laser using the
master equation in Section II. The physical significance of each term in the
master equation is elaborated and related to the quantities of interests (in
Section III) such as two-photon correlation and Duan's \cite{Duan}
entanglement measure. In Section IV, we show the importance of laser phase
for acquiring entanglement. In Section V, the steady state solutions for the
photon numbers and correlation between photon pairs are given. We show that
the laser phase provides a useful knob for controlling entanglement. We then
use the results to derive a condition for entanglement in the double Raman
scheme. By using proper values of cavity damping and laser parameters based
on analysis of the condition, we obtain macroscopic number of entangled
photon pairs in the steady state. We also analyze the double resonant Raman
scheme but find no entanglement.

\section{Physics of Two-Photon Laser}

For simplicity, we consider single atom localized in the doubly resonant
cavity. Using the usual approach \cite{QO} we obtain the master equation
\begin{equation*}
\frac{d}{dt}\hat{\rho}=[C_{\text{loss1}}(\hat{a}_{1}\hat{\rho}\hat{a}%
_{1}^{\dagger }-\hat{\rho}\hat{a}_{1}^{\dagger }\hat{a}_{1})+C_{\text{gain1}%
}(\hat{a}_{1}^{\dagger }\hat{\rho}\hat{a}_{1}-\hat{a}_{1}\hat{a}%
_{1}^{\dagger }\hat{\rho})
\end{equation*}
\begin{equation*}
+C_{\text{loss2}}(\hat{a}_{2}\hat{\rho}\hat{a}_{2}^{\dagger }-\hat{a}%
_{2}^{\dagger }\hat{a}_{2}\hat{\rho})+C_{\text{gain2}}(\hat{a}_{2}^{\dagger }%
\hat{\rho}\hat{a}_{2}-\hat{\rho}\hat{a}_{2}\hat{a}_{2}^{\dagger })+
\end{equation*}
\begin{equation}
e^{-i\varphi _{t}}(C_{\text{1}}\hat{a}_{2}\hat{\rho}\hat{a}_{1}-C_{\text{2}}%
\hat{\rho}\hat{a}_{1}\hat{a}_{2}+\ C_{\text{3}}\hat{a}_{1}\hat{\rho}\hat{a}%
_{2}-C_{\text{4}}\hat{a}_{1}\hat{a}_{2}\hat{\rho})]\text{+adj.,}
\label{master}
\end{equation}
$\allowbreak $with the effective phase $\varphi _{t}$. For double Raman
scheme, $\varphi _{t}=\varphi _{p}+\varphi _{c}-(\varphi _{s}+\varphi _{a})$%
. The phases $\varphi _{\alpha }=k_{\alpha }z+\phi _{\alpha }$ of the lasers
depend on both the position $z$ of the atom and the controllable absolute
phases $\phi _{\alpha }$ of the lasers. So, $\phi _{s}=\phi _{a}=0$. Since $%
k_{p}+k_{c}-k_{s}-k_{a}=0$, the effective phase becomes $\varphi _{t}=\phi
_{p}+\phi _{c}=\phi $. The explicit expressions for $C_{\text{lossj}}$, $C_{%
\text{gainj}}$ and $C_{\text{k}}$ (where j$=$1,2 and k$=$1,2,3,4) are given
in Appendix \ref{coefficients}.

Equation (\ref{master}) already includes the cavity damping Liouvillean $L%
\hat{\rho}=-\sum\limits_{j=1,2}\kappa _{j}\{\hat{a}_{j}^{\dagger }\hat{a}_{j}%
\hat{\rho}+\hat{\rho}\hat{a}_{j}^{\dagger }\hat{a}_{j}-2\hat{a}_{j}\hat{\rho}%
\hat{a}_{j}^{\dagger }\}$. We find that the relation holds
\begin{equation}
C_{\text{1}}+C_{\text{3}}=C_{\text{2}}+C_{\text{4}}\text{.}
\label{C relation}
\end{equation}
Note that Eq. (\ref{master}) generalizes the master equation for the cascade
scheme \cite{cascade master} in which $C_{\text{3}}=C_{\text{gain2}}=0$, and
$C_{\text{lossj}}=\kappa _{j}$.

The $C_{\text{gainj}}$ are due to the emissions processes of the atom in the
excited levels and Raman process via the laser fields which provide gain. On
the other hand, the $C_{\text{lossj}}$ are due to cavity dissipation $\kappa
_{j}$ and absorption processes of the atom in the ground levels which create
loss. The $C_{\text{k}}$ coefficients correspond to squeezing. Each term
gives the coherence between $n_{j}$ and $n_{j}\pm 1$ such that the
difference between the total photon number in the bra and in the ket is
always $2$. Figure \ref{REDlaserscheme}b illustrates in two-dimensional
photon number space the essence of each diagonal term in Eq. (\ref{master}).

The consequence of this relation is that for large number of photons $%
n_{j}>>1$, the coherences due to $\hat{a}_{2}\hat{\rho}\hat{a}_{1},\hat{\rho}%
\hat{a}_{1}\hat{a}_{2},\hat{a}_{1}\hat{\rho}\hat{a}_{2},\hat{a}_{1}\hat{a}%
_{2}\hat{\rho}$ and their adjoint are approximately equal. Hence, the
contribution of the off-diagonal terms vanish and the master equation
reduces to the rate equation. Since the off-diagonal terms give rise to
entanglement (as we show below), we can understand that there will be no
entanglement for very large $n_{j}$.

\begin{figure}[tbp]
\center\epsfxsize=8.5cm\epsffile{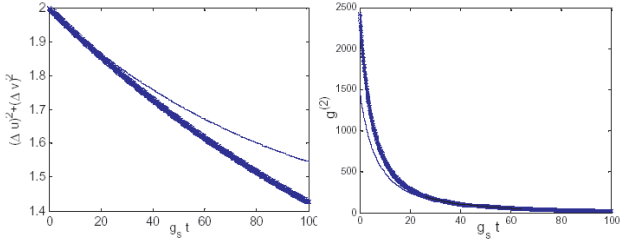}
\caption{The ${\left( {\Delta \hat{u}}\right) ^{2}+\left( {\Delta \hat{v}}%
\right) ^{2}}$ and $g^{(2)}(t)$ vary in as similar manner as a function of
time with decoherence $\protect\gamma _{bc}=0.6\protect\gamma _{ac}$ (thin
line) and without decoherence $\protect\gamma _{bc}=0$ (thick line).
Parameters used are $\protect\kappa _{s}=\protect\kappa _{a}=0.001g_{s}$, $%
\Omega _{p}=2g_{s},\Delta _{p}=40g_{s},\Omega _{c}=25g_{s},\Delta _{c}=0$.
We have assumed $g_{a}=g_{s},$ with $\protect\gamma _{ac}=\protect\gamma
_{dc}=\protect\gamma _{ab}=\protect\gamma _{db}=\protect\gamma $.}
\label{REDg2-Duan}
\end{figure}

\section{Relation between entanglement and two-photon correlation}

Two-photon correlation for \emph{Raman emission doublet} (RED) (large
detuning and weak pump) version of the double Raman scheme for single atom
\cite{HH} and extended medium \cite{paper I},\cite{Harris} show nonclassical
properties such as antibunching and violation of Cauchy-Schwarz inequality.
It is useful to show how the nonclassical correlation relates to
entanglement. The normalized two-photon correlation at zero time delay is

\begin{eqnarray}
g^{(2)}(t) &\doteq &\frac{|\langle \hat{a}_{2}\hat{a}_{1}\rangle |^{2}}{%
\langle \hat{a}_{2}^{\mathbf{\dagger }}\hat{a}_{2}\rangle \langle \hat{a}%
_{1}^{\mathbf{\dagger }}\hat{a}_{1}\rangle }+1\text{,} \\
|\langle \hat{a}_{2}\hat{a}_{1}\rangle | &=&\sqrt{\bar{n}_{1}\bar{n}%
_{2}(g^{(2)}(t)-1)}\text{.}
\end{eqnarray}
Thus, the $g^{(2)}(t)$ does not provide phase $\phi _{21}$ information of
the correlation $\langle \hat{a}_{2}\hat{a}_{1}\rangle $. We introduce the
phase via
\begin{equation}
\langle \hat{a}_{2}\hat{a}_{1}\rangle =|\langle \hat{a}_{2}\hat{a}%
_{1}\rangle |e^{i\phi _{21}}\text{.}  \label{<a2a1>}
\end{equation}
Hence the Duan's parameter $D(t)=\hspace{-1cm}{\left( {\Delta \hat{u}}%
\right) ^{2}+\left( {\Delta \hat{v}}\right) ^{2}}$ can be rewritten as${}$%
\begin{eqnarray}
D(t) &=&2[1+\bar{n}_{1}+\bar{n}_{2}+2\sqrt{\bar{n}_{1}\bar{n}%
_{2}(g^{(2)}(t)-1)}\cos \phi _{21}  \notag \\
&&-|\langle \hat{a}_{2}\rangle |^{2}-|\langle \hat{a}_{1}\rangle
|^{2}-\langle \hat{a}_{2}\rangle \langle \hat{a}_{1}\rangle -\langle \hat{a}%
_{2}^{\mathbf{\dagger }}\rangle \langle \hat{a}_{1}^{\mathbf{\dagger }%
}\rangle ]\text{.}  \label{criteria vs phi}
\end{eqnarray}
The terms in the second line would be$\ -|\alpha _{1}|^{2}-|\alpha
_{2}|^{2}-(\alpha _{1}\alpha _{2}+\alpha _{1}^{\ast }\alpha _{2}^{\ast })$
for an input coherent state. If $g^{(2)}(t)=1$ we can have entanglement that
is independent of the phase $\phi _{21}$
\begin{equation}
D(t)=2[1-(\alpha _{1}\alpha _{2}+\alpha _{1}^{\ast }\alpha _{2}^{\ast })]%
\text{.}
\end{equation}
We shall consider the vacuum state, in which they vanish. Clearly, the
presence of inseparability or entanglement is entirely determined by the
phase $\phi _{21}$ in Eq. (\ref{criteria vs phi}). We now find the knob to
control entanglement, i.e. $\cos \phi _{21}$ must be negative or $\pi
/2<\phi _{21}<3\pi /2$.

The condition for inseparability or entanglement $\hspace{-1cm}{0<\left( {%
\Delta \hat{u}}\right) ^{2}+\left( {\Delta \hat{v}}\right) ^{2}<2}$ can be
re-expressed in terms of the phase and the two-photon correlation
\begin{equation}
-\frac{\bar{n}_{1}+\bar{n}_{2}+1}{2\sqrt{\bar{n}_{1}\bar{n}_{2}(g^{(2)}-1)}}%
<\cos \phi _{21}<-\frac{\bar{n}_{1}+\bar{n}_{2}}{2\sqrt{\bar{n}_{1}\bar{n}%
_{2}(g^{(2)}-1)}}
\end{equation}
where the lower limit corresponds to maximum entanglement. The midpoint
value $\cos \phi _{21}=-\frac{\bar{n}_{1}+\bar{n}_{2}+1/2}{2\sqrt{\bar{n}_{1}%
\bar{n}_{2}(g^{(2)}-1)}}$ gives $D(t)=1$. When $\bar{n}_{1}=\bar{n}_{2}$, we
have $-\frac{1+1/2\bar{n}}{\sqrt{g^{(2)}-1}}<\cos \phi _{21}<-\frac{1}{\sqrt{%
g^{(2)}-1}}$. Note that for photon antibunching $g^{(2)}\gtrsim 1$ the
entanglement window for $\phi _{21}$ can be quite large, provided $\bar{n}%
_{1},\bar{n}_{2}$ are small too.

In the limit of large two-photon correlation (photon bunching) $g^{(2)}>>1$
and large photon numbers $\bar{n}_{1},\bar{n}_{2}>>1$ the range for
entanglement becomes quite restrictive. Here, $\cos \phi _{21}\simeq -\frac{1%
}{\sqrt{g^{(2)}-1}}$ becomes very small in magnitude (but negative) and from
Eq. (\ref{criteria vs phi}), $D(t)\lesssim 2$ the entanglement is small.
This explains the results in Fig. \ref{REDg2-Duan} where large transient
correlation is accompanied by small entanglement.

In the long time limit Fig. \ref{REDg2-Duan} shows that the correlation
vanishes (corresponds to antibunching) and there is entanglement, $D<<2$.
Thus, both the antibunching and entanglement are compatible quantum
mechanical properties since they manifest in the same way. Duan's
entanglement increases with time while the correlation decreases with time,
thus both do not vary in the same way. This clearly shows that correlation
should be discerned from entanglement. However, as expected, the decoherence
$\gamma _{bc}$ tends to reduce the degree of entanglement and the magnitude
of correlation.

\section{Laser Phase for Entanglement}

Here, we show by using simple example from the resonant cascde work of Han
\textsl{et. al.} \cite{Han} that the \emph{nonzero phase} of the paired
correlation $\langle \hat{a}_{1}\hat{a}_{2}\rangle $ is necessary for
entanglement. Let us analyze the transient equation (written in their
notations with zero laser phase)

\begin{equation*}
\frac{d\langle \hat{a}_{1}\hat{a}_{2}\rangle }{dt}=-\langle \hat{a}_{1}\hat{a%
}_{2}\rangle (\beta _{22}^{\ast }-\beta _{11}+\kappa _{2}+\kappa _{1})
\end{equation*}

\begin{equation}
-\beta _{21}^{\ast }(\langle \hat{a}_{1}^{\dagger }\hat{a}_{1}\rangle
+1)+\beta _{12}\langle \hat{a}_{2}^{\dagger }\hat{a}_{2}\rangle \text{.}
\end{equation}
The coefficients for resonant case are such that: $\beta _{11},\beta _{22}$
are real while $\beta _{12}=i\alpha _{12}$ and $\beta _{21}=i\alpha _{21}$
are purely imaginary. Clearly we have imaginary value for
\begin{equation*}
\langle \hat{a}_{1}\hat{a}_{2}\rangle (t)=i\int_{0}^{t}e^{-K(t-t^{\prime
})}\{\alpha _{12}\langle \hat{a}_{2}^{\dagger }\hat{a}_{2}\rangle (t^{\prime
})+
\end{equation*}
\begin{equation}
\alpha _{21}(\langle \hat{a}_{1}^{\dagger }\hat{a}_{1}\rangle (t^{\prime
})+1)\}dt^{\prime }\simeq iX
\end{equation}
where $K=\beta _{22}-\beta _{11}+\kappa _{2}+\kappa _{1}$ and $X$ is real
whose expression is not important for the present discussion.

For initial condition $\langle \hat{a}_{j}(0)\rangle =0$, the Duan's
criteria
\begin{equation}
D=2[1+\langle \hat{a}_{1}^{\mathbf{\dagger }}\hat{a}_{1}\rangle +\langle
\hat{a}_{2}^{\mathbf{\dagger }}\hat{a}_{2}\rangle +\langle \hat{a}_{1}\hat{a}%
_{2}\rangle +\langle \hat{a}_{1}^{\mathbf{\dagger }}\hat{a}_{2}^{\mathbf{%
\dagger }}\rangle ]  \label{D vacuum}
\end{equation}
clearly shows $D=2+2\{\langle \hat{a}_{1}^{\mathbf{\dagger }}\hat{a}%
_{1}\rangle +\langle \hat{a}_{2}^{\mathbf{\dagger }}\hat{a}_{2}\rangle
+iX-iX\}>2$, there is no entanglement.

For finite phase $\phi $ associated to the pump laser, the correlation $%
\langle \hat{a}_{1}\hat{a}_{2}\rangle $ becomes $\langle \hat{a}_{1}\hat{a}%
_{2}\rangle e^{i\phi }$ \ but the photon numbers are not affected. The
parameter becomes
\begin{equation}
D{=}2(1+\langle \hat{a}_{1}^{\mathbf{\dagger }}\hat{a}_{1}\rangle +\langle
\hat{a}_{2}^{\mathbf{\dagger }}\hat{a}_{2}\rangle +2X\sin \phi )
\end{equation}
which gives maximum entanglement when $\phi =-\pi /2$ or $3\pi /2$, and no
entanglement when $\phi =0$.

\section{Steady State Entanglement}

The master equation (\ref{master}) is linear and do not include saturation.
One might wonder whether steady state solutions. We find that there are
steady state solutions when the photon numbers $\bar{n}_{j}$ do not increase
indefinitely. Parameters that give non steady state solutions manifest as
negative value of $D$ and should be disregarded. The study of entanglement
via nonlinear theory will be published elsewhere.

In the case of initial vacuum, only $\frac{d\bar{n}_{1}}{dt},\frac{d\bar{n}%
_{2}}{dt},\frac{d\langle \hat{a}_{1}\hat{a}_{2}\rangle }{dt},\frac{d\langle
\hat{a}_{1}^{\dagger }\hat{a}_{2}^{\dagger }\rangle }{dt}$ are sufficient to
compute Duan's entanglement parameter, where $\bar{n}_{j}=\langle \hat{a}%
_{j}^{\dagger }\hat{a}_{j}\rangle $, $j=1,2$. From the master equation (\ref
{master}), we obtain the coupled equations 
\begin{eqnarray}
\frac{d\bar{n}_{1}}{dt} &=&\bar{n}_{1}K_{1}+e^{-i\phi }(C_{\text{1}}-C_{%
\text{2}})\langle \hat{a}_{1}\hat{a}_{2}\rangle +  \notag \\
&&e^{i\phi }(C_{\text{1}}^{\ast }-C_{\text{2}}^{\ast })\langle \hat{a}%
_{1}^{\dagger }\hat{a}_{2}^{\dagger }\rangle +C_{\text{gain1}}+C_{\text{gain1%
}}^{\ast }\text{,}  \label{dn1/dt}
\end{eqnarray}
\begin{eqnarray}
\frac{d\bar{n}_{2}}{dt} &=&\bar{n}_{2}K_{2}+e^{-i\phi }(C_{\text{3}}-C_{%
\text{2}})\langle \hat{a}_{1}\hat{a}_{2}\rangle +  \notag \\
&&e^{i\phi }(C_{\text{3}}^{\ast }-C_{\text{2}}^{\ast })\langle \hat{a}%
_{1}^{\dagger }\hat{a}_{2}^{\dagger }\rangle +C_{\text{gain2}}+C_{\text{gain2%
}}^{\ast }\text{,}  \label{dn2/dt}
\end{eqnarray}
\begin{equation}
(\frac{d}{dt}-K_{12})\langle \hat{a}_{1}\hat{a}_{2}\rangle =e^{i\phi }[\bar{n%
}_{1}(C_{\text{3}}^{\ast }-C_{\text{2}}^{\ast })+\bar{n}_{2}(C_{\text{1}%
}^{\ast }-C_{\text{2}}^{\ast })-C_{\text{2}}^{\ast }]  \label{da1a2/dt}
\end{equation}
where the gain coefficients/loss are
\begin{eqnarray}
K_{j} &=&C_{\text{gainj}}+C_{\text{gainj}}^{\ast }-(C_{\text{lossj}}+C_{%
\text{lossj}}^{\ast }) \\
K_{12} &=&C_{\text{gain2}}+C_{\text{gain1}}^{\ast }-(C_{\text{loss2}}+C_{%
\text{loss1}}^{\ast })\text{.}
\end{eqnarray}

The steady state solution for the correlation is $\langle \hat{a}_{1}\hat{a}%
_{2}\rangle =Ee^{i\phi }$ where
\begin{eqnarray*}
E &=&[-C_{32}^{\ast }(K_{2}K_{12}^{\ast }+C_{12}^{\ast
}C_{32}-C_{12}C_{32}^{\ast })(C_{\text{gain1}}+C_{\text{gain1}}^{\ast }) \\
&&-C_{12}^{\ast }(K_{1}K_{12}^{\ast }+C_{12}C_{32}^{\ast }-C_{12}^{\ast
}C_{32})(C_{\text{gain2}}+C_{\text{gain2}}^{\ast }) \\
&&+C_{2}^{\ast }(K_{1}C_{12}C_{32}^{\ast }+K_{2}C_{32}C_{12}^{\ast
})-C_{2}^{\ast }K_{1}K_{2}K_{12}^{\ast }
\end{eqnarray*}
\begin{equation}
-C_{2}C_{32}^{\ast }C_{12}^{\ast }(K_{1}+K_{2})]\frac{1}{M}
\end{equation}
where
\begin{eqnarray}
M &=&(K_{1}K_{12}^{\ast }+K_{2}K_{12})\left( C_{12}^{\ast }C_{32}\right) +%
\text{c.c.}  \notag \\
&&-\left( C_{12}C_{32}^{\ast }-C_{12}^{\ast }C_{32}\right)
^{2}-K_{1}K_{2}K_{12}K_{12}^{\ast }\text{.}
\end{eqnarray}
The steady state solutions for the photon numbers are
\begin{eqnarray}
\bar{n}_{1} &=&(C_{\text{gain1}}+C_{\text{gain1}}^{\ast })\frac{%
K_{2}K_{12}K_{12}^{\ast }-(C_{12}^{\ast }C_{32}K_{12}^{\ast }+\text{c.c.})}{M%
}  \notag \\
&&+(C_{\text{gain2}}+C_{\text{gain2}}^{\ast })C_{12}^{\ast }C_{12}\frac{%
K_{12}^{\ast }+K_{12}}{M}  \notag \\
&&+C_{\text{2}}^{\ast }C_{12}\frac{K_{2}K_{12}^{\ast }+(C_{12}^{\ast }C_{32}-%
\text{c.c.})}{M}  \notag \\
&&+C_{\text{2}}C_{12}^{\ast }\frac{K_{2}K_{12}+(C_{12}C_{32}^{\ast }-\text{%
c.c.})}{M}\text{,}  \label{n1 st}
\end{eqnarray}
\begin{eqnarray}
\bar{n}_{2} &=&(C_{\text{gain2}}+C_{\text{gain2}}^{\ast })\frac{%
K_{1}K_{12}K_{12}^{\ast }-(C_{32}^{\ast }C_{12}K_{12}^{\ast }+\text{c.c.})}{M%
}  \notag \\
&&+(C_{\text{gain1}}+C_{\text{gain1}}^{\ast })C_{32}^{\ast }C_{32}\frac{%
K_{12}^{\ast }+K_{12}}{M}  \notag \\
&&+C_{\text{2}}^{\ast }C_{32}\frac{K_{1}K_{12}^{\ast }+(C_{12}C_{32}^{\ast }-%
\text{c.c.})}{M}  \notag \\
&&+C_{\text{2}}C_{32}^{\ast }\frac{K_{1}K_{12}+(C_{12}^{\ast }C_{32}-\text{%
c.c.})}{M}\text{.}  \label{n2 st}
\end{eqnarray}

From Eq. (\ref{D vacuum}), the necessary condition for entanglement is $%
Ee^{i\phi }+E^{\ast }e^{-i\phi }<-(\bar{n}_{1}+\bar{n}_{2})$. If $E$ is real
positive there would be no entanglement in the region $\cos \phi >0$.
Entanglement is still possible even if $\phi =0$ provided real\{$E$\}$<0$.
Thus, the phase $\phi $ is not necessary for entanglement, but it provides
an \emph{extra knob} for controlling entanglement.

Let us search for entanglement conditions for the limiting cases of
Raman-EIT (REIT) scheme which produces nonclassically correlated photon
pairs, and the double resonant Raman (DRR) scheme.

\subsection{Raman-EIT regime}

For this scheme $\Omega _{c},\Delta _{c}(=\Delta )>>\Omega _{p},\gamma _{x}$
($x=ab,ac,db,dc,bc$ indices for decoherence rates) and $\Delta =0$. Thus, we
have $p_{ba}=\frac{-i\Omega _{c}^{\ast }}{\gamma _{ab}}(p_{bb}-p_{aa})%
\rightarrow 0$, $p_{cd}=\frac{-\Omega _{p}}{\Delta }=p_{dc}$. It follows
from Appendix A that the only finite coefficients in the master equation are
\begin{equation}
C_{\text{loss1}}\simeq \kappa _{s},C_{\text{loss2}}\simeq \kappa _{a},C_{%
\text{2}}\simeq -C_{\text{4}}\simeq i\Xi \text{,}
\end{equation}
\begin{equation}
K_{1}=-2\kappa _{s},K_{2}=-2\kappa _{a},K_{12}=-(\kappa _{s}+\kappa _{a})
\label{K for REIT}
\end{equation}
where
\begin{equation}
\Xi =g_{a}g_{s}\frac{\Omega _{p}\Omega _{c}}{\Delta (\gamma _{ac}\gamma
_{bc}+\Omega _{c}^{2})}\text{.}  \label{CAS}
\end{equation}

Here, we also have which are used to to write the steady state solutions
\begin{equation}
n_{1}=\Xi ^{2}\frac{\kappa _{a}}{(\kappa _{s}+\kappa _{a})[\kappa _{a}\kappa
_{s}-\Xi ^{2}]}\text{,}  \label{n1 st REIT}
\end{equation}
\begin{equation}
n_{2}=\Xi ^{2}\frac{\kappa _{s}}{(\kappa _{s}+\kappa _{a})[\kappa _{a}\kappa
_{s}-\Xi ^{2}]}\text{,}  \label{n2 st REIT}
\end{equation}
\begin{equation}
\langle \hat{a}_{1}\hat{a}_{2}\rangle =e^{i\theta _{t}}i\Xi \frac{\kappa
_{a}\kappa _{s}}{(\kappa _{s}+\kappa _{a})[\kappa _{a}\kappa _{s}-\Xi ^{2}]}%
\text{.}  \label{a12 st REIT}
\end{equation}
The entanglement criteria can be rewritten as
\begin{equation}
\Xi \frac{\Xi -\frac{\kappa _{a}\kappa _{s}}{(\kappa _{s}+\kappa _{a})}2\sin
\theta _{t}}{\kappa _{a}\kappa _{s}-\Xi ^{2}}<0\text{.}
\label{condition REIT}
\end{equation}
For negative detuning $\Xi <0$, there are two possibilities: a) if $\kappa
_{a}\kappa _{s}<\Xi ^{2}$ entanglement occurs in the region $\frac{\kappa
_{a}\kappa _{s}}{(\kappa _{s}+\kappa _{a})}2\sin \theta _{t}>\Xi $, b) if $%
\kappa _{a}\kappa _{s}>\Xi ^{2}$ we have entanglement in $\frac{\kappa
_{a}\kappa _{s}}{(\kappa _{s}+\kappa _{a})}2\sin \theta _{t}<\Xi $.
Similarly, for positive detuning $\Xi >0$: a) if $\kappa _{a}\kappa _{s}<\Xi
^{2}$ then we need $\Xi >\frac{\kappa _{a}\kappa _{s}}{(\kappa _{s}+\kappa
_{a})}2\sin \theta _{t}$ and b) if $\kappa _{a}\kappa _{s}>\Xi ^{2}$ then we
need $\Xi <\frac{\kappa _{a}\kappa _{s}}{(\kappa _{s}+\kappa _{a})}2\sin
\theta _{t}$.

To obtain large entanglement, we tune the cavity damping such that $|\kappa
_{a}\kappa _{s}-C_{2}^{\ast }C_{2}|$ is small and $\sin $ $\theta _{t}\sim 1$%
. Note that the \emph{sign} of the detuning $\Delta $ in Eq. (\ref{CAS}) is
important for entanglement generation. We can arrange the signs of these
quantities such that Eq. (\ref{condition REIT}) is satisfied.

\begin{figure*}[tbp]
\center\epsfxsize=15cm\epsffile{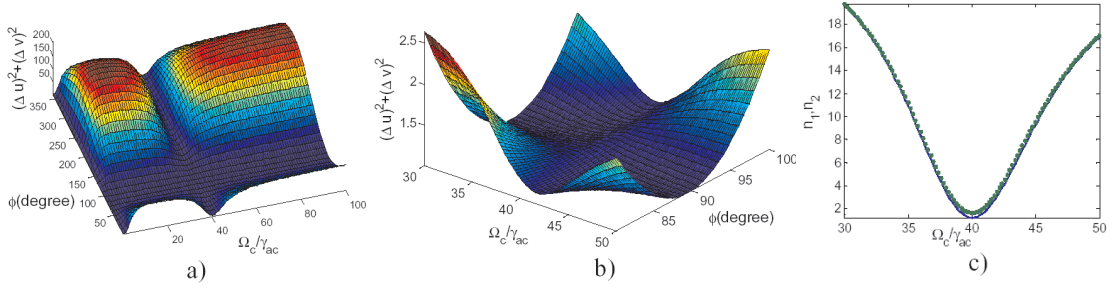}
\caption{Entanglement for Raman-EIT scheme in a) wide view, b) magnified
view of focused region (square box) in a), and c) mean photon numbers $\bar{n%
}_{1}$(solid line) and $\bar{n}_{2}$(dots), where $\Omega _{p}=\protect\gamma
_{ac},\Delta _{p}=40\protect\gamma _{ac},\Delta _{c}=0$. Cavity damping
values $\protect\kappa _{a}=\protect\kappa _{s}=\protect\sqrt{1.01\left(
C_{2}^{\ast }C_{2}\right) }$ ensures a minimum value of the denominator of
the condition in Eq. (\ref{condition REIT}). We have assumed $g_{a}=g_{s},%
\protect\gamma _{bc}=0$ with $\protect\gamma _{ac}=\protect\gamma _{dc}=%
\protect\gamma _{ab}=\protect\gamma _{db}=\protect\gamma $. }
\label{REDlaserSTRamanEIT}
\end{figure*}

Figure \ref{REDlaserSTRamanEIT} is plotted using $\kappa _{a}=\kappa _{s}=%
\sqrt{1.01\left( C_{2}^{\ast }C_{2}\right) }$ and $\Delta =40\gamma _{ac}$
for $\theta _{t}\sim 90^{o}$. We find that the region $\Omega _{c}\sim
\Delta $ gives a large entanglement, but the photon numbers are minimum.
This prevents having a steady state macroscopic entanglement. We verify that
if we change to a negative detuning $\Delta =-40\gamma _{ac}$ there is no
entanglement.

The region of maximum entanglement occur around $\phi =90^{0}$. Entanglement
can occur at a wide range of large $\Omega _{c}$. However, the photon
numbers $\bar{n}_{j}$ decrease with the increase of $\Omega _{c}$.

\begin{figure}[tbp]
\center\epsfxsize=8.7cm\epsffile{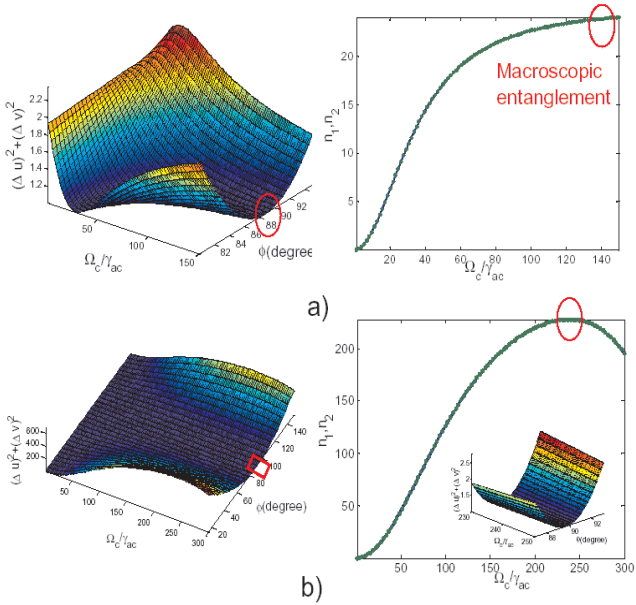}
\caption{Macroscopic entanglement for a) larger $\Omega
_{p}=10\protect\gamma _{ac},\Delta _{p}=400\protect\gamma _{ac}$
cavity damping and other
parameters the same as Fig. \ref{REDlaserSTRamanEIT}, b) $\Omega _{p}=10%
\protect\gamma _{ac},\Delta _{p}=400\protect\gamma _{ac}$ but cavity damping
$\protect\kappa _{a}=\protect\kappa _{s}=\protect\sqrt{1.001\left(
C_{2}^{\ast }C_{2}\right) }$. This gives larger mean photon numbers and
macroscopic entangled photon pairs.}
\label{REDlaserSTRamanEITmacro}
\end{figure}

Figure \ref{REDlaserSTRamanEITmacro} shows that it is possible to obtain a
continuous bright source of entangled photons. We realize that the number of
nonclassical photon pairs in REIT scheme is limited by the weak pump field.
Thus, by increasing the pump field we can generate more Stokes photons (Fig.
\ref{REDlaserSTRamanEITmacro}a). At the same time, the detuning is increased
as well to ensures that the scheme remain in the nonclassical Raman-EIT
regime. By applying the derived condition Eq. (\ref{condition REIT}) we
further obtained a larger number of entangled photon pairs (Fig. \ref
{REDlaserSTRamanEITmacro}b).

\subsection{Resonant Case}

It seems that steady state entanglement in the double resonant Raman case ($%
\Omega _{c}=\Omega _{p},\Delta _{c}=\Delta _{p}=0$) is hardly possible. In
the following, we investigate this analytically. Here, we find $%
C_{ac,ac},C_{ac,bd},C_{bd,ac},C_{bd,bd}$ \ are real and positive while $%
C_{ac,ad},C_{ac,bc},C_{bd,ad},C_{bd,bc}$ are purely imaginary (positive or
negative). Since $p_{cd}=-i|p_{cd}|,p_{ba}=-i|p_{ba}|$ all $C_{\text{1}},C_{%
\text{2}},C_{\text{3}},C_{\text{4}},$ $C_{\text{lossj}}$ and $C_{\text{gainj}%
}$ are real and could be negative. So, $K_{j}=2C_{\text{gainj}}-2C_{\text{%
lossj}}$ and $K_{12}=\frac{1}{2}(K_{1}+K_{2})$.

For \emph{symmetric} system, $\Omega _{p}\simeq \Omega _{c}$ we find $%
p_{ab}=-p_{ba}=p_{dc}=-p_{cd}$, $p_{cc}\simeq p_{bb}$ and $p_{aa}\simeq
p_{dd}$ \cite{pop range}. Then, $C_{ac,ac}=C_{bd,bd}$ and $%
C_{ac,bd}=C_{bd,ac}$. If we take $T_{ac}=T_{dc}=T_{ab}=T_{db}=\gamma $
(spontaneous decay rate) with $T_{bc}=\gamma _{bc}$\ and $T_{ad}=2\gamma $
we further have $C_{bd,ad}=-C_{ac,ad},C_{bd,bc}=-C_{ac,bc}$.

The steady state solutions for DRR scheme can be written as
\begin{equation}
\bar{n}_{1}=\bar{n}_{2}=\frac{C_{\text{gain}}(C_{\text{gain}}-C_{\text{loss}%
})+\frac{1}{2}C_{2}C_{12}}{C_{12}^{2}-(C_{\text{gain}}-C_{\text{loss}})^{2}}%
\text{,}  \label{nj st DRR}
\end{equation}
\begin{equation}
\langle \hat{a}_{1}\hat{a}_{2}\rangle =-e^{i\theta _{t}}\left( \frac{C_{1}C_{%
\text{gain}}-\frac{1}{2}C_{2}(C_{\text{gain}}+C_{\text{loss}})}{%
C_{12}^{2}-(C_{\text{gain}}-C_{\text{loss}})^{2}}\right)  \label{a12 st DRR}
\end{equation}
and the entanglement condition with initial vacuum as
\begin{equation}
\bar{n}_{1}+\bar{n}_{2}<2\xi \cos \theta _{t}  \label{condition DRR}
\end{equation}
where $\xi $ is the term in the bracket $(...)$.

In order to determine whether Eq. (\ref{condition DRR}) can be met we
consider a simpler case where $\gamma _{bc}=0$. From Appendix B, we have $C_{%
\text{loss}}-C_{\text{gain}}=\kappa $ and $C_{12}=\frac{g^{2}}{\gamma }%
(p_{cc}-p_{aa})$, $C_{1}=\frac{g^{2}}{\gamma }p_{aa}$, $C_{2}=\frac{g^{2}}{%
\gamma }(2p_{aa}-p_{cc})$, $C_{\text{loss}}=(\frac{g^{2}}{2\gamma }%
p_{bb}+\kappa )$, $C_{\text{gain}}=\frac{g^{2}}{2\gamma }p_{cc}$. These
results are used to rewrite Eqs. (\ref{nj st DRR}) and (\ref{a12 st DRR}) as
\begin{equation}
\bar{n}_{j}=\frac{g^{2}}{2\gamma }\frac{-p_{cc}\kappa +\frac{g^{2}}{\gamma }%
(2p_{aa}-p_{cc})(p_{cc}-p_{aa})}{[\frac{g^{2}}{\gamma }(p_{cc}-p_{aa})]^{2}-%
\kappa ^{2}}  \label{nj st nodec DRR}
\end{equation}
\begin{equation}
\langle \hat{a}_{1}\hat{a}_{2}\rangle =e^{i\theta _{t}}\frac{g^{2}}{2\gamma }%
\frac{(2p_{aa}-p_{cc})(\frac{g^{2}}{2\gamma }2p_{cc}+\kappa )-\frac{g^{2}}{%
\gamma }p_{aa}p_{cc}}{[\frac{g^{2}}{\gamma }(p_{cc}-p_{aa})]^{2}-\kappa ^{2}}
\label{a12 st nodec DRR}
\end{equation}

For strong field, $p_{cc}\simeq p_{aa}=0.25$. The steady solutions become $%
\bar{n}_{1}=\bar{n}_{2}=\frac{g^{2}}{8\gamma \kappa }$, $\langle \hat{a}_{1}%
\hat{a}_{2}\rangle =-\frac{g^{2}}{8\gamma \kappa }e^{i\theta _{t}}$ and $%
D=2(1+\frac{g^{2}}{2\gamma \kappa }\sin ^{2}\frac{1}{2}\theta _{t})$, i.e.
no entanglement.

For weak field, $p_{cc}\simeq 0.5,p_{aa}\simeq 0$. The steady solutions are $%
\bar{n}_{1}=\bar{n}_{2}=\frac{g^{2}}{4\gamma }\frac{1}{\kappa -\frac{g^{2}}{%
2\gamma }}$ , $\langle \hat{a}_{1}\hat{a}_{2}\rangle =\frac{g^{2}}{4\gamma }%
\frac{e^{i\theta _{t}}}{\kappa -\frac{g^{2}}{2\gamma }}$ with $\kappa >\frac{%
g^{2}}{2\gamma }$ and hence $D=2(1+\frac{2g^{2}}{2\gamma \kappa -g^{2}}\sin
^{2}\frac{1}{2}\theta _{t})$, again no entanglement. In the weak field
regime, the cavity damping has to be sufficiently large to ensure the
existence of steady state solutions, i.e. $\kappa >g^{2}/2\gamma $. If the
cavity damping is small $\kappa <g^{2}/2\gamma $ , regions with negative
values of $D$ and $\bar{n}_{j}$ would appear which reflect the non-steady
state regime.

Thus, we have shown that there is \emph{no} steady state entanglement for
DRR scheme in both weak and strong fields regimes. This is compatible with
its classical two-photon correlation $G^{(2)}$\cite{G2 for DRR}. However,
the REIT photon pairs, which are nonclassically correlated, are also
entangled in the steady state.

\section{Conclusion}

We have shown that two-photon laser can produce a continuous source of
entangled photon pairs based on the steady state solutions and an
entanglement criteria. We have obtained a relationship between entanglement
and two-photon correlation, and find that both do not vary with time in the
same manner. We have derived a condition for steady state entanglement in
the Raman-EIT (REIT) schemes and showed that steady state macroscopic
entanglement is possible. We find that a large steady state entanglement
occurs at the expense of smaller photon numbers. We showed that the double
resonant Raman (DRR) does not generate steady state entangled photon pairs
for any laser parameters.

\appendix

\section{Coefficients for double Raman scheme}

\label{coefficients} The coefficients in Eq. \ref{master} for double Raman
scheme are
\begin{eqnarray}
C_{\text{loss1}} &=&|g_{s}|^{2}(C_{bd,ad}p_{ab}+C_{bd,bd}p_{bb})+\kappa _{s},
\\
C_{\text{gain1}} &=&|g_{s}|^{2}\{C_{bd,bd}p_{dd}+C_{bd,bc}p_{dc}\}, \\
C_{\text{loss2}} &=&|g_{a}|^{2}(C_{ac,ac}p_{cc}+C_{ac,ad}p_{cd})+\kappa _{a},
\\
C_{\text{gain2}} &=&|g_{a}|^{2}\{C_{ac,ac}p_{aa}+C_{ac,bc}p_{ba}\},
\end{eqnarray}
\begin{eqnarray}
J_{1} &=&C_{bd,ac}p_{cc}+C_{ac,bd}^{\ast }p_{dd}+(C_{bd,ad}+C_{ac,bc}^{\ast
})p_{cd} \\
J_{2} &=&C_{bd,ac}p_{aa}+C_{ac,bd}^{\ast
}p_{dd}+C_{bd,bc}p_{ba}+C_{ac,bc}^{\ast }p_{cd} \\
J_{3} &=&C_{bd,ac}p_{aa}+C_{ac,bd}^{\ast }p_{bb}+(C_{bd,bc}+C_{ac,ad}^{\ast
})p_{ba} \\
J_{4} &=&C_{bd,ac}p_{cc}+C_{ac,bd}^{\ast
}p_{bb}+C_{bd,ad}p_{cd}+C_{ac,ad}^{\ast }p_{ba}
\end{eqnarray}
where $J_{k}=\frac{C_{k}}{g_{a}g_{s}}$, $g_{a},g_{s}$ are atom-field
coupling strengths, $C_{\alpha \beta ,\gamma \delta }$ ($\alpha ,\beta
,\gamma ,\delta =a,b,c,d$) are complex coefficients that depend on
decoherence rates $\gamma _{\alpha \beta }$, laser detunings $\Delta _{p}$ ,
$\Delta _{c}$ and Rabi frequencies $\Omega _{p}$, $\Omega _{c}$. The $%
p_{\alpha \alpha },p_{ab},p_{cd}$ ($\alpha =a,b,c,d$) are steady state
populations and coherences.

\begin{eqnarray}
C_{ac,ac} &=&\frac{T_{ad}^{\ast }T_{bc}^{\ast }T_{db}+I_{p}T_{ad}^{\ast
}+I_{c}T_{bc}^{\ast }}{Z} \\
C_{ac,ad} &=&-i\Omega _{p}\frac{T_{bc}^{\ast }T_{db}+I_{p}-I_{c}}{Z}
\end{eqnarray}

\begin{eqnarray}
C_{ac,bc} &=&-i\Omega _{c}\frac{-T_{ad}^{\ast }T_{db}+I_{p}-I_{c}}{Z} \\
C_{ac,bd} &=&\Omega _{c}\Omega _{p}\frac{T_{bc}^{\ast }+T_{ad}^{\ast }}{Z}
\end{eqnarray}

\begin{eqnarray}
C_{bd,ac} &=&\Omega _{p}\Omega _{c}\frac{T_{bc}^{\ast }+T_{ad}^{\ast }}{Z} \\
C_{bd,ad} &=&-i\Omega _{c}\frac{-T_{ac}^{\ast }T_{bc}^{\ast }+I_{p}-I_{c}}{Z}
\end{eqnarray}

\begin{eqnarray}
C_{bd,bc} &=&-i\Omega _{p}\frac{T_{ac}^{\ast }T_{ad}^{\ast }+I_{p}-I_{c}}{Z}
\\
C_{bd,bd} &=&\frac{T_{ac}^{\ast }T_{ad}^{\ast }T_{bc}^{\ast
}+I_{p}T_{bc}^{\ast }+I_{c}T_{ad}^{\ast }}{Z}
\end{eqnarray}

\begin{eqnarray}
Z &=&T_{ac}^{\ast }T_{ad}^{\ast }T_{bc}^{\ast }T_{db}+I_{p}T_{ac}^{\ast
}T_{ad}^{\ast }+I_{p}T_{bc}^{\ast }T_{db}  \notag \\
&&+I_{c}T_{ac}^{\ast }T_{bc}^{\ast }+I_{c}T_{ad}^{\ast
}T_{db}+(I_{p}-I_{c})^{2}
\end{eqnarray}

where $I_{p}=\Omega _{p}^{2}$ and $I_{c}=\Omega _{c}^{2}$.

\section{Coefficients for RRD scheme}

From Appendix A, we obtain the coefficients for RRD scheme:
\begin{eqnarray}
C_{\text{1}} &=&g_{a}g_{s}\frac{\Omega ^{2}}{Z}2\{T_{bc}p_{cc}+T_{ad}p_{aa}\}
\\
C_{\text{2}} &=&g_{a}g_{s}\frac{\Omega ^{2}}{Z}%
2\{T_{bc}p_{aa}+T_{ad}(2p_{aa}-p_{cc})\}
\end{eqnarray}
\begin{eqnarray}
C_{12} &=&C_{\text{1}}-C_{\text{2}}=C_{\text{3}}-C_{\text{2}}  \notag \\
&=&g_{a}g_{s}\Omega ^{2}\frac{T_{bc}+T_{ad}}{Z}2(p_{cc}-p_{aa})  \label{C12}
\end{eqnarray}
\begin{equation}
Z=\gamma \{T_{ad}T_{bc}\gamma +2I(T_{ad}+T_{bc})\}\text{.}  \label{Z}
\end{equation}
Taking $\kappa _{s}=\kappa _{s}=\kappa $ we also have $C_{\text{loss1}}=C_{%
\text{loss2}}$, $C_{\text{gain1}}=C_{\text{gain2}}$ and $%
K_{2}=K_{1}=K_{12}=2(C_{\text{gain}}-C_{\text{loss}})$, so
\begin{eqnarray}
C_{\text{loss}} &=&|g_{s}|^{2}(\frac{IT_{bc}}{Z}p_{aa}+T_{ad}\frac{TT_{bc}+I%
}{Z}p_{bb})+\kappa ,  \label{Closs} \\
C_{\text{gain}} &=&|g_{s}|^{2}\{T_{bc}\frac{TT_{ad}+I}{Z}p_{dd}+\frac{IT_{ad}%
}{Z}p_{cc}\},  \label{Cgain}
\end{eqnarray}



\begin{thebibliography}{99}
\bibitem{pop range}  When the fields are weak, the populations are
distributed only in the lower levels, so $p_{cc},p_{bb}=0.5$ and $%
p_{aa},p_{dd}=0$. When the fields are very large, all levels are equally
populated; $p_{aa}=p_{bb}=p_{cc}=p_{dd}=0.25$. So, the ranges $0.25\leqslant
p_{cc},p_{bb}\leqslant 0.5$ and $0\leqslant p_{aa},p_{dd}\leqslant 0.25$
apply for the double resonant Raman scheme (DRR).
\bibitem{G2 for DRR}  Results of two-photon correlation for double resonant
Raman will be published as a sequel to ref. \cite{paper I}.
\end{thebibliography}
\end{document}